\pdfoutput=1
\documentclass[sigconf]{acmart}
\renewcommand\footnotetextcopyrightpermission[1]{} 
\usepackage{enumitem}

\begin{document}
\title{Fine-tune BERT for E-commerce Non-Default Search Ranking}

\author{Yunjiang Jiang\textsuperscript{\textsection}}
\email{yunjiang.jiang@jd.com}
\affiliation{JD.com, Mountain View, 94043} 

\author{Yue Shang\textsuperscript{\textsection}}
\email{yue.shang@jd.com}
\affiliation{JD.com, Mountain View, 94043}

\author{Hongwei Shen}
\affiliation{JD.com, Mountain View, 94043}

\author{Wenyun Yang}
\affiliation{JD.com, Mountain View, 94043}

\author{Yun Xiao}
\affiliation{JD.com, Mountain View, 94043}


\begin{abstract}
  The quality of non-default ranking on e-commerce platforms, such as based on ascending item price or descending historical sales volume, often suffers from acute relevance problems, since the irrelevant items are much easier to be exposed at the top of the ranking results. In this work, we propose a two-stage ranking scheme, which first recalls wide range of candidate items through refined query/title keyword matching, and then classifies the recalled items using BERT-Large fine-tuned on human label data. We also implemented parallel prediction on multiple GPU hosts and a C++ tokenization custom op of Tensorflow. In this data challenge, our model won the 1st place in the supervised phase (based on overall F1 score) and 2nd place in the final phase (based on average per query F1 score). 
\end{abstract}
\keywords{E-Commerce Search, Information Retrieval, Deep Learning}

\maketitle

\begingroup\renewcommand\thefootnote{\textsection}
\footnotetext{These authors contributed equally to this research.}
\endgroup

\section{Introduction}
Learning relevance between query and item is one of the most fundamental problems in e-Commerce search ranking. But it has unique requirements compared with traditional semantic matching problems. For e-Commerce search, unlike web search, item side textual information typically has limited length compared with traditional document retrieval. 

Within the e-Commerce relevance ranking domain, non-default ranking is an important but challenging area. On the one hand, the search results are sorted by other dimensions, such as price, comments, or freshness, which may inadvertently expose the irrelevant items to users, thereby hurting user experiences. For example, the query ``guitar" incorrectly retrieves the accessory item ``5Pcs B-2 Tone replace parts metal guitar strings lines" on a major US e-Commerce platform. On the other hand, failure to recall the relevant items is easily detected by the sellers and can result in profit loss of both the platform and sellers, as well as adverse user experience. Thus, this high accuracy recall task\cite{trotman2018high}\cite{degenhardt2019ecom} is key to a healthy ecosystem on e-Commerce platforms.

BERT (Bidirectional Encoder Representations from Transformers) is a popular and powerful pre-trained multi-layer transformer model released by Google AI in 2018\cite{devlin2019bert}. It achieved state of the art results on multiple benchmark natural language challenges, such as the Stanford Question and Answer Dataset. More details on how to use the pre-trained model can be found on its official github page\cite{bert2018github}.

In this work, we investigate ways to introduce BERT into e-Commerce non-default search ranking. Our contributions are as follows:
\begin{itemize}[noitemsep, topsep=0pt]
    \item Using a two-stage ranking scheme to incorporate BERT into high-accuracy recall search ranking task.
    \item Implement parallel prediction on multiple GPU hosts and a C++-based BERT tokenizer via Tensorflow custom op.
    \item Experiment and compare different BERT fine-tuning schemes and features.
\end{itemize}
\section{Related Work}

The task of e-Commerce search ranking has been extensively studied in the past. Traditional methods that predate neural nets include RankNet~\cite{L2RPoint}, RankSVM~\cite{ranksvm}, GBRank~\cite{gbrank}), and list-wise models like AdaRank~\cite{AdaRank} and LambdaMart~\cite{LambdaMart}; see \cite{l2rSurvey} for a complete survey. 

With the explosion of deep learning in the recent decade, we see largely a dichotomy of new approaches to ranking. In the first category, embeddings of the query and item are learned first in an unsupervised manner, which are subsequently fed as features in a supervised phase. This has been explored in \cite{van2016learning}, \cite{yu2014latent}, and \cite{ai2017learning}. The other approach, pioneered by the DSSM\cite{huang2013learning} model initially conceived in 2013, learns the ranking task end-to-end. Several noteworthy follow-up works include DRMM\cite{guo2016deep}, Duet\cite{duet}, and DeepRank\cite{deep_rank}.

Literature on non-default search ranking methods on any platform is very sparse. We were not able to find anything useful in the e-Commerce setting. However the methods listed above can be easily adapted to a binary classification task for the purpose of filtering irrelevant results.

\section{Methods}
\begin{figure}
    \centering
    \includegraphics[width=0.4\textwidth]{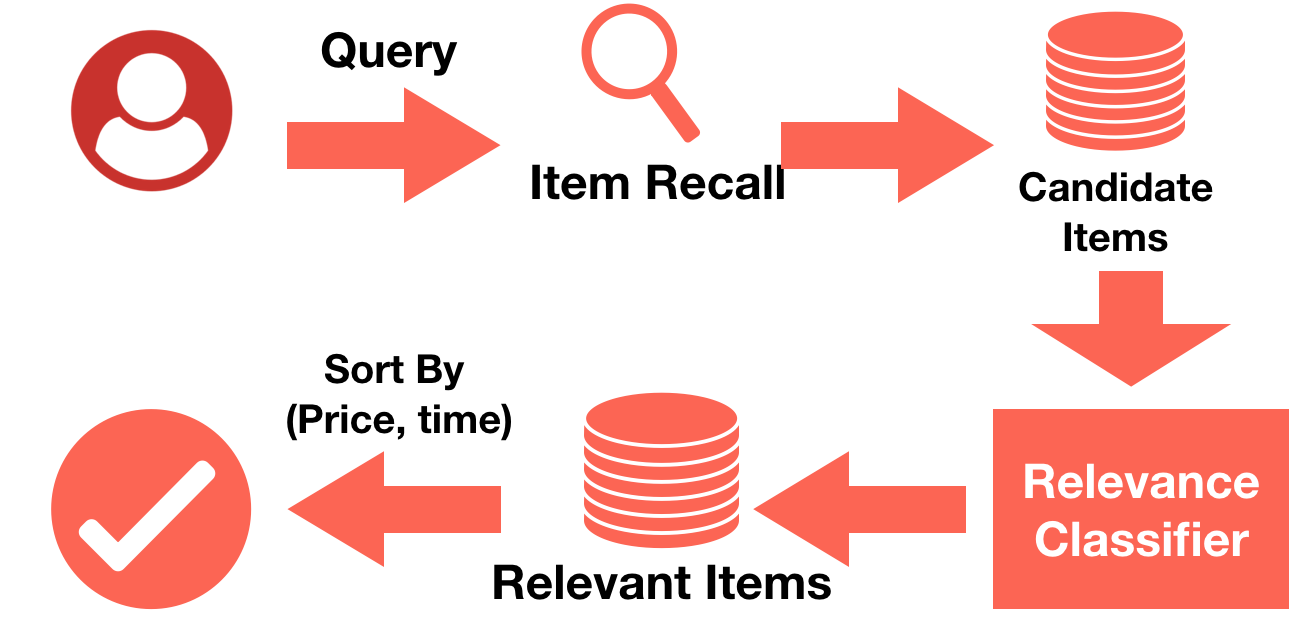}
    \caption{General process of Two-stage document retrieval model}
    \label{fig:search_infra}
\end{figure}

In this part, we will briefly introduce methods and models we tried in this task. First, we describe how we clean the data, and how we conducted coarse-grained item recall based on semantic term matching. Then we will focus on different fine-tuning methods to achieve high precision recall tradeoff. The general process is as illustrated in Figure \ref{fig:search_infra}

\subsection{Coarse-grained Item Recall}
We applied some widely used NLP techniques to perform text data pre-processing for term matching based item recall, specifically, a pipeline with tokenizer, stopword removal, and stemmer. And we also used synonyms to expand queries, by adding the synonyms to the original query tokens. 

\begin{itemize}
    \item Tokenization: queries and titles were both tokenized by discarding special characters and splitting along whitespaces.
    \item Stopword removal: adopted stopwords set from NLTK\footnote{NLTK: https://www.nltk.org/}. 
    \item Stemmer: use word root to represent words in different forms. Here we use SnowballStemmer\cite{snowball}.
    \item Synonyms Expansion: we applied a synonyms dictionary to include more related items, e.g. ``usmc'' to ``united states marine corps''. 
\end{itemize}
According to the corpus offered by data challenge, there are 899k documents and 150 queries, producing 150 millions of query/item pairs. After the coarse-grained item recall process, we eliminated the candidate pairs to approximately 5 millions.

\subsection{Fine-tune BERT Model}
BERT (Bidirectional Encoder Representations from Transformers)\cite{devlin2019bert}, as a pre-training language model, achieved the state-of-the-art results on many tasks, such as sentiment analysis\cite{sun2019utilizing} and question answering\cite{dua2019drop}. The model architecture is based on multi-layer Transformer encoder\cite{vaswani2017attention}. BERT's base model consists of 12 layers, 768 hidden units, and 12 attention heads. It is pre-trained on large text corpus such as Wikipedia, using combined losses from masked language model and next sentence prediction. 

BERT takes an input sequence with a fix length (128 tokens by default) and outputs the representation of sequence. The input sequence can represent one or two pieces of text, using a special token [SEP] to separate. And the first token of the sequence is specially designed as [CLS], which is a label embedding of the input sequence. 

In our fine-tuning step, the input data was a triple of (query, item title, label). We found that rather than the default 128 token length, the maximum total token size of query and title combined is only 64. This dramatically reduced the memory requirement of training, allowing us to use bigger batch sizes for better performance. This was especially critical when using the BERT-large pretrained model, even on an Nvidia V100 GPU card which is one of the best in the market as of early 2019. We managed to use a batch size of 28 on the latter device.

We also split the supervised data set into training and eval of roughly 90 to 10 ratio. Initially we made sure that the two data sets had disjoint sets of queries, as a gold standard for avoiding over-fitting. Eventually however we split examples under each query with the same ratio between training and eval, to ensure better query coverage during training, since some queries were severely under-represented in the overall labeled data set.

We loaded the pre-trained model released in \cite{bert2018github} at the beginning of training, and concatenated query and title as input sequence to fine tune BERT in a binary text classification task. The pooled representation of final layer, $\textbf{h}$, given by the embedding of the special token [CLS], was passed into a simple logistic regression classifier to learn the probability that the title represented a relevant document for the query.  
\begin{equation}
    P(\text{relevance}|\textbf{h}) = \text{softmax}(W\textbf{h}),
\end{equation}
where $W$ is the parameter to learn in the fine-tuning step, whose dimension is $\text{num\_hidden\_size}*2$.

\subsection{Second-Phase Fine-Tuning}
Once the BERT model converged in the held-out eval data set, according to batch AUC, we experimented with secondary fine-tuning on top of the second-to-last softmax layer, using a small multi-layer perceptron. We tried linear, and 2-layer MLP approaches. While the linear model took a long time to converge, and did not significantly exceed the eval accuracy of the baseline BERT model, the 2-layer MLP did improve significantly by about 0.6\%. This was sufficiently motivating to try scoring the full 150m (query, item) pairs with it. However the performance on the hidden evaluation set was slightly worse than vanilla BERT model. By ensembling with the vanilla BERT scores, however, this improved our final evaluation F1 by roughly another percentage point.

\subsection{Ensemble}
We also tried to ensemble the prediction results of different BERT models by simply averaging the scores. 
\begin{equation}
    \bar{s}=\frac{1}{N}\sum_{s_i\in S^{(N)}}{s_i}
\end{equation}

\subsection{Other methods}
\subsubsection{Price and Breadcrumb Features}
Given that the task was about price ascending ranked results, it was natural to add price to the list of features. In addition, the breadcrumb (hierarchical category) feature looked very promising, especially since it was directly visible to the user. 

We experimented with adding these two features into the training data, by simply combining them with the title into a single ``answer'' feature, using arbitrary separator such as the bar ascii character ``|'', which is discarded anyway during tokenization. This increased the maximum token length from 64 to 78 in the supervised training data, which is still far shorter than the default 128 token length, so with BERT-Large Whole Word we can still use batch size of 24 on a V100.

The resulting eval accuracy improved by about 1 point, which we did not consider high enough to warrant inclusion in the final model, due to the added data complexity.

\subsubsection{ImageNet Models}
For many items, title or more general text features alone may not be the most informative or authoritative in relevance judgment. Thus we resorted to item images, whose urls were provided by the competition data set. It took us roughly 1 day to crawl all 900k item images. A small fraction of those were PNG or TIF, but the majority were of JPEG format. 

We used standard tf image decoder to convert the subset of those images in the training set into tf example format, along with the query viewed as one of 151 categories. The label can be any integer from -150 to 150, where a negative label stands for a negative example, and the label 150 stands for a positive example belonging to a class outside the 150 queries (otherwise the positive labels are 0-based).

Then we used the tutorial ImageNet model Inception V3 in tf.slim to train a multi-class classification model. 
For negative examples, we modified the conventional softmax layer by the statistical consideration that a negative label means the example belongs to one of the remaining 150 categories (including the label 150 which again means not in any of the 150 given queries).

The resulting eval F1 score from the image model turned out to be far worse than those of the BERT fine-tuned model, around 0.75\% at best; BERT achieved well over 0.8\%, see Table~\ref{tab:eval-ai-result}. Ensembling with the BERT model did not significantly degrade the results of the latter model, but also did not yield measurable improvement.

\section{Scalable Prediction}
While the total number of query, item pairs is over 150m, our recall selection procedure lowered it to about 5m. This was still a very large number of examples to score by an expensive model like BERT, especially if we wanted to iterate fast and try ensembles of multiple BERT fine-tuned models. Therefore we spent significant amount of time investigating how to maximize our GPU resources. Essentially we had about 5 GPU machines, each equipped with 8 Ti1080 Nvidia GPUs.

\subsection{Parallel scoring on a single GPU host}
It is relatively straightforward to use all GPUs on a single machine to do model scoring: simply divide the test file into 8 equal pieces, and spawn 8 jobs simultaneously to score each one. We found that very small batch sizes can slow down prediction job significantly, but once the batch size reaches above 128, bigger batches did not make much difference, even though the GPU memory consumption went up linearly.

\subsection{Parallel scoring on multiple GPU hosts}
Next we tried to utilize multiple machines to run BERT scoring simultaneously. We first investigated the naive idea of making GPUs on other machines visible in the OS environment of the chief machine. This turned out to be impossible, since GPU cards require installation to communicate with the host CPU.

A brute force approach would be to shard the data locally, then send equal fractions of the shards to other machines and manually launch prediction jobs there. 

To eliminate the manual step and bookkeeping challenge, we mounted the host machine onto the 5 GPU machines via NFS, then during scoring, we issue a master command on the chief host that would run ssh command into each of the 5 worker hosts, each of which in turn launches 8 BERT prediction jobs on data stored on the chief host, using python/tensorflow binaries on the chief host, and storing the predicted scores on the chief host as well. In other words, the worker hosts are used only for 1. their CPUs for data ingestion and processing and 2. GPU cards for model forward pass and nothing else.

\begin{figure}[h]
    \centering
    \includegraphics[width=0.4\textwidth]{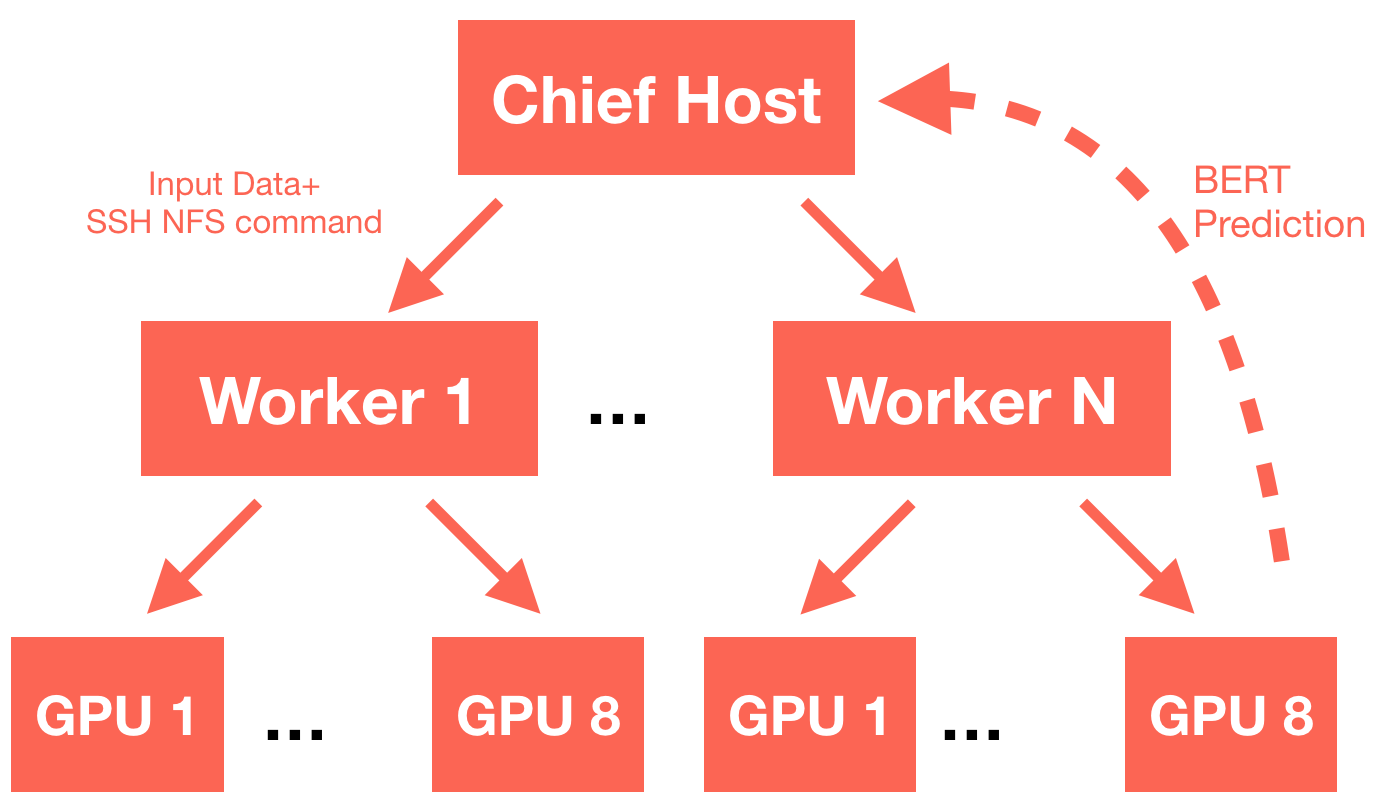}
    \caption{Parallel prediction}
    \label{fig:parallel}
\end{figure}

\subsection{C++ based streaming tokenization}
Another slow-down factor in the original BERT implementation is that the input examples need to be converted from tab-separated format to tf.Example format, before training, evaluation, and model scoring.

While it is possible to parallel process the conversion job using python multiprocessing module, it created extra bookkeeping hassle. In addition, different models (such as BERT versus XLNET, or BERT Whole-Word versus BERT Large) could require multiple copies of the converted tf.Example binary files in order to accommodate different tokenization methods. 

To eliminate this complication, we developed tensorflow-based C++ tokenization custom op to convert query and titles into their respective BERT prescribed tokens. The resulting op must take the vocab file as an input. This differs from more conventional tokenizers, such as the unigram/bigram tokenizer, which are vocab file agnostic.

In addition, since BERT combines the query and title tokens into a single fixed length token id list, with special tokens such as [CLS] and [SEP], we need a way to dynamically stitch the query and title tokens together, in order to conform with the pre-training methodology. This we did in python, using a series of numpy/tensorflow operations.

\begin{figure}[h]
    \centering
    \includegraphics[width=0.15\textwidth]{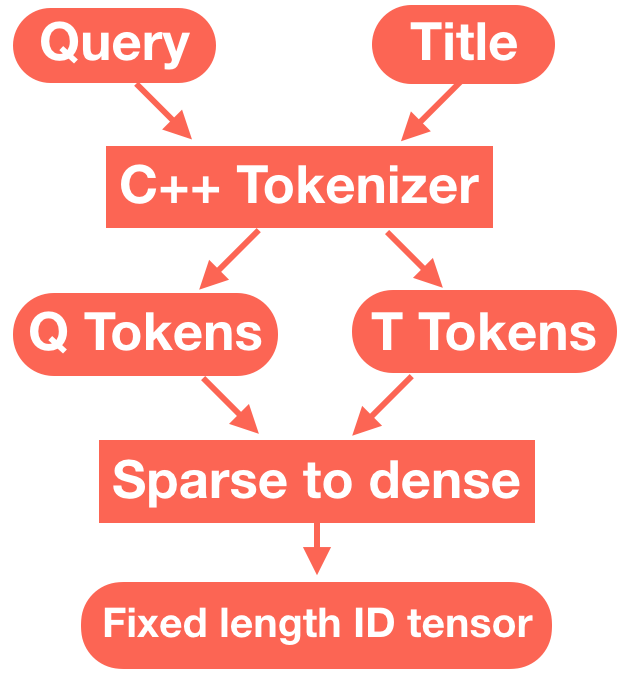}
    \caption{C++ Streaming Tokenizer}
    \label{fig:tokenizer}
\end{figure}

\section{Experiments and Result Analysis}
We list several results from models we submitted to evalAI\cite{EvalAI} system as shown in Table \ref{tab:eval-ai-result}. The model tuned from Whole Word Masking BERT-Large model(bert.wwm) has a slight improvement(0.37\%) compare to BERT-Base model. Adding a multilayer perceptron of 32 hidden units(bert.wwm.mlp32) to the latter gives the highest F1 score(0.8293) among single models. Ensemble models boost results further. We ensembled 5 models and obtained F1-score of 0.8451. During the competition, we received 1st place in the supervised phase and 2nd place of the final phase. 
 
\begin{table}[h]
\begin{tabular}{llll}
\hline
Model               & Precision & Recall & F1 Score \\ \hline
bert.base.uncased   & 0.8168    & 0.8314 & 0.8241   \\
bert.wwm            & 0.7509    & 0.9207 & 0.8272   \\
bert.wwm.mix\_title & 0.8183    & 0.8045 & 0.8114   \\ 
bert.wwm.mlp32      & 0.8086    & 0.8510 & 0.8293   \\
bert.ensemble\_3     & 0.7655    & 0.9149 & 0.8336   \\
bert.ensemble\_4     & 0.7744    & 0.9152 & 0.8390   \\
\hline
\end{tabular}
\caption{Submissions in supervised phase from EvalAI}
\label{tab:eval-ai-result}
\end{table}

\section{Conclusion}
We presented in detail our solution for the SIGIR 2019 `High accuracy recall task' data challenge, based primarily on query and item title text features. The main workhorse for the classification task is a fine-tuned BERT model trained on relevance-labeled query/item pairs provided by the competition host. To reduce the total number of scoring pairs from 150m (cartesian product of all queries and items) down to about 5m, we employed NLP techniques such as stemming, tokenization, and synonym expansion to filter the potential candidate set. To further speed up BERT prediction, we implemented a parallel scoring framework on a GPU cluster based on NFS. In addition, we reported exploratory results leveraging other features provided, including item images, prices, and hierarchical categories (breadcrumbs). In the future, we hope to explore other related methodologies, such as BERT distillation and implicit feedback learning.

\section{Acknowledgement}
We thank Rui Li (previously at JD.com) for introducing us to the present competition, as it is closely related to our own work.

\bibliographystyle{ACM-Reference-Format}


\begin{thebibliography}{0}


\ifx \showCODEN    \undefined \def \showCODEN     #1{\unskip}     \fi
\ifx \showDOI      \undefined \def \showDOI       #1{#1}\fi
\ifx \showISBNx    \undefined \def \showISBNx     #1{\unskip}     \fi
\ifx \showISBNxiii \undefined \def \showISBNxiii  #1{\unskip}     \fi
\ifx \showISSN     \undefined \def \showISSN      #1{\unskip}     \fi
\ifx \showLCCN     \undefined \def \showLCCN      #1{\unskip}     \fi
\ifx \shownote     \undefined \def \shownote      #1{#1}          \fi
\ifx \showarticletitle \undefined \def \showarticletitle #1{#1}   \fi
\ifx \showURL      \undefined \def \showURL       {\relax}        \fi
\providecommand\bibfield[2]{#2}
\providecommand\bibinfo[2]{#2}
\providecommand\natexlab[1]{#1}
\providecommand\showeprint[2][]{arXiv:#2}

\end{thebibliography}


\begin{thebibliography}{00}

\bibitem{ai2017learning}
Qingyao Ai, Yongfeng Zhang, Keping Bi, Xu~Chen, and W~Bruce Croft.
\newblock Learning a hierarchical embedding model for personalized product
  search.
\newblock In {\em Proceedings of the 40th International ACM SIGIR Conference on
  Research and Development in Information Retrieval}, pages 645--654. ACM,
  2017.

%
\bibitem{L2RPoint}
Chris Burges, Tal Shaked, Erin Renshaw, Ari Lazier, Matt Deeds, Nicole
  Hamilton, and Greg Hullender.
\newblock Learning to rank using gradient descent.
\newblock In {\em Proceedings of the 22Nd International Conference on Machine
  Learning}, ICML '05, pages 89--96, New York, NY, USA, 2005. ACM.

\bibitem{LambdaMart}
Christopher J.~C. Burges.
\newblock From {RankNet} to {LambdaRank} to {LambdaMART}: An overview.
\newblock Technical report, Microsoft Research, 2010.

%
%
%
\bibitem{degenhardt2019ecom}
Jon Degenhardt, Surya Kallumadi, Utkarsh Porwal, and Andrew Trotman.
\newblock Ecom'19: The sigir 2019 workshop on ecommerce.
\newblock In {\em Proceedings of the 42nd International ACM SIGIR Conference on
  Research and Development in Information Retrieval}, pages 1421--1422. ACM,
  2019.

\bibitem{bert2018github}
Jacob Devlin.
\newblock Tensorflow code and pre-trained models for bert.
\newblock \url{https://github.com/google-research/bert}, 2018.

\bibitem{devlin2019bert}
Jacob Devlin, Ming-Wei Chang, Kenton Lee, and Kristina Toutanova.
\newblock Bert: Pre-training of deep bidirectional transformers for language
  understanding.
\newblock In {\em Proceedings of the 2019 Conference of the North American
  Chapter of the Association for Computational Linguistics: Human Language
  Technologies, Volume 1 (Long and Short Papers)}, pages 4171--4186, 2019.

\bibitem{dua2019drop}
Dheeru Dua, Yizhong Wang, Pradeep Dasigi, Gabriel Stanovsky, Sameer Singh, and
  Matt Gardner.
\newblock Drop: A reading comprehension benchmark requiring discrete reasoning
  over paragraphs.
\newblock In {\em Proceedings of the 2019 Conference of the North American
  Chapter of the Association for Computational Linguistics: Human Language
  Technologies, Volume 1 (Long and Short Papers)}, pages 2368--2378, 2019.

\bibitem{guo2016deep}
Jiafeng Guo, Yixing Fan, Qingyao Ai, and W~Bruce Croft.
\newblock A deep relevance matching model for ad-hoc retrieval.
\newblock In {\em Proceedings of the 25th ACM International on Conference on
  Information and Knowledge Management}, pages 55--64, 2016.

%

\bibitem{huang2013learning}
Po-Sen Huang, Xiaodong He, Jianfeng Gao, Li~Deng, Alex Acero, and Larry Heck.
\newblock Learning deep structured semantic models for web search using
  clickthrough data.
\newblock In {\em Proceedings of the 22nd ACM international conference on
  Information \& Knowledge Management}, pages 2333--2338, 2013.

\bibitem{ranksvm}
Thorsten Joachims.
\newblock Optimizing search engines using clickthrough data.
\newblock In {\em Proceedings of the Eighth ACM SIGKDD International Conference
  on Knowledge Discovery and Data Mining}, KDD '02, pages 133--142, 2002.


\bibitem{l2rSurvey}
Tie-Yan Liu.
\newblock Learning to rank for information retrieval.
\newblock {\em Found. Trends Inf. Retr.}, 3(3):225--331, March 2009.

%
%
\bibitem{duet}
Bhaskar Mitra, Fernando Diaz, and Nick Craswell.
\newblock Learning to match using local and distributed representations of text
  for web search.
\newblock In {\em Proceedings of the 26th International Conference on World
  Wide Web}, WWW '17, pages 1291--1299, 2017.

\bibitem{deep_rank}
Liang Pang, Yanyan Lan, Jiafeng Guo, Jun Xu, Jingfang Xu, and Xueqi Cheng.
\newblock Deeprank: A new deep architecture for relevance ranking in
  information retrieval.
\newblock In {\em Proceedings of the 2017 ACM on Conference on Information and
  Knowledge Management}, CIKM '17, pages 257--266, 2017.

\bibitem{snowball}
Martin Porter.
\newblock Snowball compiler and stemming algorithms.
\newblock \url{https://github.com/snowballstem/snowball}, 2019.

%
%
\bibitem{trotman2018high}
Andrew Trotman, Surya Kallumadi, and Jon Degenhardt.
\newblock High accuracy recall task.
\newblock 2018.

\bibitem{van2016learning}
Christophe Van~Gysel, Maarten de~Rijke, and Evangelos Kanoulas.
\newblock Learning latent vector spaces for product search.
\newblock In {\em Proceedings of the 25th ACM International on Conference on
  Information and Knowledge Management}, pages 165--174. ACM, 2016.

\bibitem{vaswani2017attention}
Ashish Vaswani, Noam Shazeer, Niki Parmar, Jakob Uszkoreit, Llion Jones,
  Aidan~N Gomez, {\L}ukasz Kaiser, and Illia Polosukhin.
\newblock Attention is all you need.
\newblock In {\em Advances in neural information processing systems}, pages
  5998--6008, 2017.

\bibitem{AdaRank}
Jun Xu and Hang Li.
\newblock Adarank: A boosting algorithm for information retrieval.
\newblock In {\em Proceedings of the 30th Annual International ACM SIGIR
  Conference on Research and Development in Information Retrieval}, SIGIR '07,
  pages 391--398, 2007.

\bibitem{EvalAI}
Deshraj Yadav, Rishabh Jain, Harsh Agrawal, Prithvijit Chattopadhyay, Taranjeet
  Singh, Akash Jain, Shiv~Baran Singh, Stefan Lee, and Dhruv Batra.
\newblock Evalai: Towards better evaluation systems for ai agents.
\newblock 2019.

%

\bibitem{yu2014latent}
Jun Yu, Sunil Mohan, Duangmanee~Pew Putthividhya, and Weng-Keen Wong.
\newblock Latent dirichlet allocation based diversified retrieval for
  e-commerce search.
\newblock In {\em Proceedings of the 7th ACM international conference on Web
  search and data mining}, pages 463--472. ACM, 2014.

\bibitem{gbrank}
Zhaohui Zheng, Keke Chen, Gordon Sun, and Hongyuan Zha.
\newblock A regression framework for learning ranking functions using relative
  relevance judgments.
\newblock In {\em Proceedings of the 30th Annual International ACM SIGIR
  Conference on Research and Development in Information Retrieval}, SIGIR '07,
  pages 287--294, 2007.

\end{thebibliography}

\end{document}